\documentclass{PoS}
\pdfoutput=1

\usepackage{amsmath}
\usepackage{amssymb}
\usepackage{graphicx}
\usepackage{dsfont}

\def\Slash#1{\setbox0=\hbox{$#1$} 
\dimen0=\wd0 
\setbox1=\hbox{/} \dimen1=\wd1 
\ifdim\dimen0>\dimen1 
\rlap{\hbox to \dimen0{\hfil/\hfil}} 
#1 
\else 
\rlap{\hbox to \dimen1{\hfil$#1$\hfil}} 
/ 
\fi}

\title{Meson and nucleon properties from Dyson-Schwinger QCD }

\ShortTitle{Meson and nucleon properties from Dyson-Schwinger QCD }

\author{\speaker{Gernot Eichmann}         \\
        Institut für Physik, Karl-Franzens-Universität Graz, A-8010 Graz, Austria\\
        E-mail: \email{ge.eichmann@uni-graz.at} }

\author{Reinhard Alkofer        \\
        Institut für Physik, Karl-Franzens-Universität Graz, A-8010 Graz, Austria\\
        E-mail: \email{reinhard.alkofer@uni-graz.at} }

\author{Andreas Krassnigg         \\
        Institut für Physik, Karl-Franzens-Universität Graz, A-8010 Graz, Austria\\
        E-mail: \email{andreas.krassnigg@uni-graz.at} }

\author{Diana Nicmorus         \\
        Institut für Physik, Karl-Franzens-Universität Graz, A-8010 Graz, Austria\\
        E-mail: \email{diana.nicmorus@uni-graz.at} }

\abstract{
        We present selected results from a calculation of meson and nucleon observables in a
        Green-function approach. A rainbow-ladder truncation of QCD's Dyson-Schwinger equations is used to
        solve the respective meson, diquark and quark-diquark Bethe-Salpeter equations.
        It allows for a simultaneous description of meson and nucleon masses and electromagnetic properties from an effective quark-gluon interaction.
        The results describe a hadronic quark core which agrees with lattice data for heavy quarks
        whereas pion-cloud effects are missing towards the chiral limit.
        The neutron's Dirac form factor is negative due to the presence of axial-vector diquark correlations.
        }

\FullConference{8th Conference Quark Confinement and the Hadron Spectrum\\
		 September 1-6, 2008\\
		 Mainz. Germany}

\begin{document}


\section{Introduction}

    It has been a longstanding challenge to describe meson and baryon properties in
    terms of QCD's elementary degrees of freedom, namely quarks and gluons.
    A quantum-field theoretical framework must address non-perturbative phenomena of QCD
    such as confinement and dynamical chiral symmetry breaking (DCSB).
    The latter is the mechanism that generates a non-perturbative constituent-quark mass scale and large dynamical hadron masses,
    and it accounts for a massless pion as the Goldstone boson in the chiral limit.
    A suitable tool to address these problems is provided by the Dyson-Schwinger equations (DSEs) of QCD
    which represent an infinite set of coupled integral equations for QCD's Green functions (see \cite{Fischer2006,Roberts:2007jh} for recent reviews).
    Combined with hadronic bound-state equations, i.e., the Bethe-Salpeter equation (BSE) for mesons and its three-body analogue for baryons,
    it offers a covariant and non-perturbative continuum approach that complements existing studies in lattice QCD, chiral effective field theories (ChEFTs) and  quark models.

    An extensive amount of meson properties has been collected in this setup during the past decade, see \cite{Maris:2006ea} for a short summary.
    Additional assumptions are needed to make the bound-state framework computationally accessible towards baryon physics.
    Owing to a strong attraction in the color-antitriplet $qq$ channel \cite{Cahill:1987qr,Hellstern:1997nv},
    correlations between two quarks may be assumed to dominate the binding of baryons.
    This simplifies the three-body equation to an effective two-body problem
    where a colorless baryon is constructed out of colored quarks and "diquarks".
    The resulting quark-diquark BSE expresses the baryon's binding through
    a quark exchange between quark and diquark which thereby swap their roles.
    The setup has been utilized to investigate the nucleon's mass and electromagnetic form factors \cite{Hellstern:1997pg,Oettel:2000jj,Oettel:2002wf,Alkofer:2004yf}
    and was recently extended to arbitrary current-quark masses \cite{Cloet2008,Cloet:2008re}.
    Observing that the diquark is bound by the same mechanism that binds quarks and antiquarks to mesons,
    a parallel effort has been made to relate meson and baryon observables through a single
    effective, current-mass dependent quark-gluon interaction \cite{Eichmann:2007nn,Eichmann2008a,Eichmann:2008ef,Nicmorus:2008vb}.
    In this paper we summarize the setup of this latter program and highlight some related results.

    An important role in the chiral structure of hadrons is played by the light pseudoscalar mesons
    which augment the hadronic 'quark core' and provide further attraction.
    Such contributions are not captured in the rainbow-ladder (RL) truncation of the Dyson-Schwinger equations which is employed herein
    and must be additionally accounted for (see, e.g., \cite{Fischer:2008wy}).
    Given that a proper implementation of chiral corrections would shift RL results in line with experiment,
    we identify those results with an inflated hadronic 'quark core'.

\section{The quark core of mesons}

    A solution of the meson BSE
    relies upon an expression for the dressed quark propagator which is obtained from solving its DSE.
    The correct implementation of spontaneous chiral symmetry breaking relates the kernels of both equations which,
    in the simplest consistent setup, is realized by a
    rainbow-ladder truncation. It is expressed by a dressed-gluon ladder exchange between quark and antiquark
    and a truncation of the quark-gluon vertex to its vector component $\sim\gamma^\mu$,
    where an effective coupling $\alpha_\text{eff}$ absorbs the
    quantities which are not explicitly solved for.
    We employ the parametrization of Refs. \cite{Maris:1999nt,Eichmann2008a}, schematically written as:
    \begin{equation}\label{mt}
        \alpha_\text{eff}(k^2,\hat{m},\omega) = c(\hat{m})\,\alpha_\text{IR}(k^2,\omega) + \alpha_\text{UV}(k^2),
    \end{equation}
    where $k^2$ is the gluon momentum, $\hat{m}$ the renormalization-point independent current-quark mass
    which enters the quark DSE, and $\omega$ a width parameter. The ultraviolet part of Eq.\,(\ref{mt}) is fixed by perturbative QCD.
    The exponential-like infrared term is modeled to provide the necessary strength to
    enable DCSB.
    Its current-mass dependence $c(\hat{m})$ was determined in \cite{Eichmann2008a} by requiring that
    $\pi$- and $\rho$-BSE solutions reproduce the following quark-core estimate for $m_\rho(m_\pi^2)$, cf. Fig.\,\ref{fig:1}:
            \begin{equation}\label{core-mrho}
                x_\rho^2 = 1 + x_\pi^4/(0.6+x_\pi^2), \;\; x_\rho = m_\rho/m_\rho^0, \;\; x_\pi = m_\pi/m_\rho^0,
            \end{equation}
    with the chiral-limit value $m_\rho^0 = 0.99$ GeV.
    Guided by lattice results and experimental data, it is based upon the assumption that beyond-RL corrections which are relevant for hadronic observables
    predominantly appear in the chiral region, partly owing to pseudoscalar meson-cloud contributions, and hence
    transcend the 'quark core' obtained in a RL truncation.
    A variation of $\omega$ by $20\%$
    has no impact on pseudoscalar and vector-meson ground-state observables;
    and, as demonstrated below, only minimally affects nucleon properties.
    Eq.\,(\ref{core-mrho}) therefore provides the only active physical input throughout our calculation
    of hadron masses and their static electromagnetic properties.

    In agreement with model expectations,
    mass-dimensionful observables ($f_\pi$, $1/r_\pi$, $\langle\bar{q}q\rangle^{1/3}$) uniformly respond to the inflated quark core for $m_\rho$;
    i.e., they are consistently overestimated by $\sim 35\%$ in the chiral limit \cite{Eichmann2008a}.
    Moreover, $f_\pi$ and $r_\pi$ tend to approach lattice results for heavier quarks (see Fig.\,\ref{fig:1}).
    This validates the notion of a pseudoscalar meson cloud which increases a hadron's charge distribution towards the chiral limit
    where it would diverge.

\section{Nucleon mass and electromagnetic properties}

\begin{figure*}
                    \begin{center}
                    \includegraphics[scale=0.88]{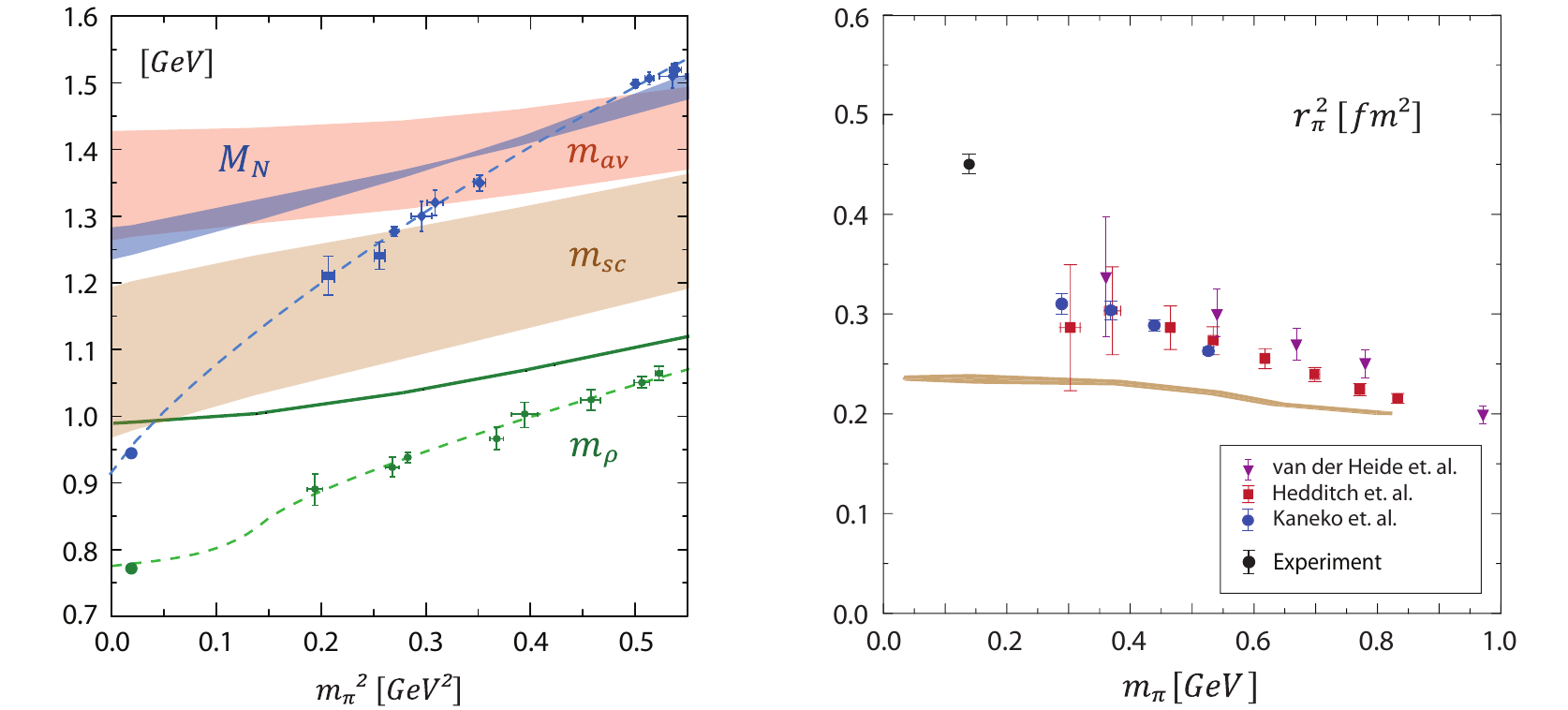}
                    \caption{ \textit{Left panel:} Evolution of the nucleon mass (\textit{thin band}) and scalar and axial-vector diquark masses (\textit{thick bands}) with
                                                  pion mass squared. The solid curve for $m_\rho$ represents the input of Eq.\,(\protect\ref{core-mrho}); the bands
                                                  denote the sensitivity to a variation of $\omega$. We compare to a selection of
                                                  lattice data and their chiral extra\-polations (\textit{dashed lines}) for $m_\rho$ and $M_N$.
                             \textit{Right panel:} Evolution of the squared pion charge radius with $m_\pi$ compared to lattice results.
                             Dots denote the experimental values.
                             (Figures adapted from Refs.~\cite{Eichmann:2008ef} and ~\cite{Eichmann2008a}; see references therein.)
                               }\label{fig:1}
                    \end{center}
        \end{figure*}

    Among the lightest diquark correlations which enter the nucleon's quark-diquark BSE and hence provide the dominant attraction
    are scalar and axial-vector diquarks. They are determined self-consistently from their diquark BSEs which are expressed by
    the same iterated gluon exchange that binds quarks and antiquarks to mesons.
    It is remarkable that, despite the large $\omega$-dependence of the unobservable diquark masses, the resulting mass of the nucleon is
    very weakly dependent on this parameter (cf. Fig.\,\ref{fig:1}), owing to cancellations in scalar and axial-vector diquark channels.
    The chosen quark core of the $\rho$-meson which leaves room for chiral corrections, Eq.\,(\ref{core-mrho}), consistently
    leads to an increased nucleon core mass, with $M_N = 1.26(2)$ GeV at the
    physical $u/d$-quark mass value \cite{Eichmann:2008ef}. Comparable values have been quoted, e.g., by ChEFT
    \cite{Young:2002c}. They suggest that the rainbow-ladder induced quark-diquark setup, despite omissions
    related to further truncations in the baryon sector, contributes a sizeable amount
    to the nucleon's quark core.

       \begin{figure*}[b]
                    \begin{center}
                    \includegraphics[scale=0.10]{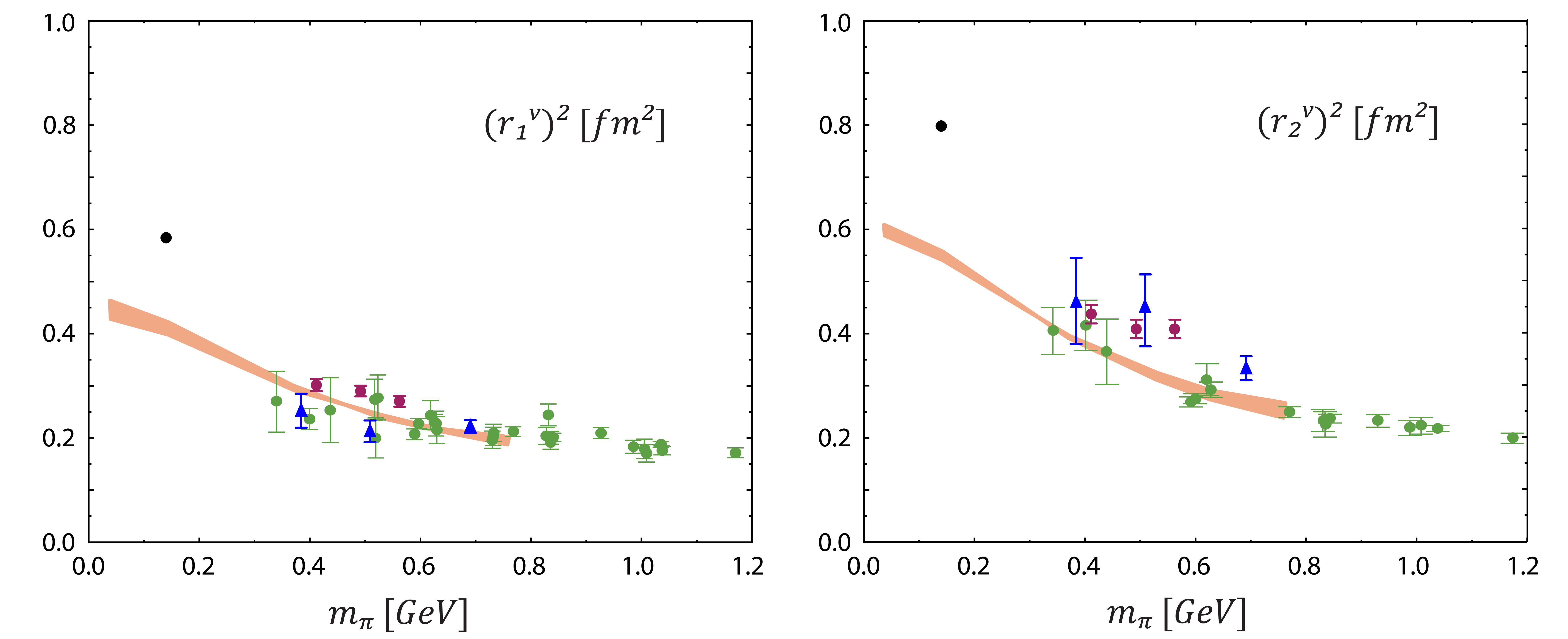}
                    \caption{Isovector radii corresponding to the Dirac and Pauli form factors $F_{1,2}^v=F_{1,2}^p-F_{1,2}^n$ compared to lattice results \cite{Alexandrou:2006ru,Gockeler:2007ir}.
                             Dots denote the experimental values.}\label{fig:2}
                    \end{center}
        \end{figure*}

    The study of electromagnetic form factors at finite photon momentum transfer requires know\-ledge of how the photon couples to the quark and diquark ingredients.
    The relevant diagrams in the quark-diquark model were derived in \cite{Oettel:1999gc} and extended to the present framework in \cite{Eichmann:2007nn}.
    Only the components transverse to the photon momentum, i.e. those not constrained by current conservation, determine the physical form factor content.
    The available information on those parts at larger $Q^2$ is limited within the scope of the current approach; however they are mandatory to enable a realistic $Q^2$-evolution of the form factors
    and the proton's form factor ratio $\mu_p G_E^p/G_M^p$ \cite{Eichmann:2008ef}.
    The respective contribution to the quark-photon vertex is known from its inhomogeneous BSE solution \cite{Maris:1999bh} and includes a $\rho$-meson pole
    which amounts to $\sim 50\%$ of both pion and nucleon squared charge radii. The nucleon's Dirac and Pauli radii $r_1$ and $r_2$ follow a similar pattern as $r_\pi$, cf. Fig.\,\ref{fig:2}:
    they are weakly dependent on $\omega$ and agree with lattice data at larger quark masses where the 'quark core' becomes the baryon.

    A natural feature of a quark-diquark model is the negativity of $F_1^n(Q^2)$.
    The presence of an axial-vector $dd$ diquark correlation centers the $d$-quark in the neutron and induces $r_1^u>r_1^d$ \cite{Eichmann:2008ef}.
    Our result for the scalar-diquark contribution to $(r_1^n)^2=(2/3)((r_1^u)^2-(r_1^d)^2)$ at the light-quark mass is $0.00(1)$ fm$^2$; adding axial-axial and scalar-axial correlations
    yields $0.11(1)$ fm$^2$. This is large compared to the experimental value $(r_1^n)^2=0.01$ fm$^2$ and indicative of
    further destructive interference with meson-cloud corrections in the axial-vector diquark channel.


   \section{Summary}

        The Green-function approach provides a consistent description of meson and baryon
        properties, which in a rainbow-ladder truncation is limited to the hadronic quark
        core.
        The framework can in principle be applied to any hadron, both its ground state and
        excitations.
        Pion-cloud effects become important towards the chiral limit, and the introduced setup
        may be used as a convenient starting point for implementing those corrections.

    \medskip

  \textbf{Acknowledgements}.
        We would like to thank C.\,D.~Roberts and I.~Cloet for helpful discussions.
        This work has been supported by the Austrian Science Fund FWF under Projects No.~P20592-N16, P20496-N16,
	    and Doctoral Program No.~W1203 as well as by the BMBF grant 06DA267.

\end{document}